\documentclass[aps,prl,twocolumn,amsmath,amssymb,footinbib,showpacs,longbibliography,superscriptaddress]{revtex4-1}

\usepackage[english]{babel}
\usepackage{latexsym}
\usepackage{graphics}
\usepackage{extarrows}
\usepackage{subfigure}
\usepackage{epsfig}
\usepackage{color}
\usepackage{hyperref}
\usepackage{braket} 
\usepackage[T1]{fontenc}
\usepackage[latin9]{inputenc}
\usepackage{amsthm,amssymb,amsfonts}
\usepackage{graphicx}
\usepackage{esint}
\usepackage{cleveref}
\usepackage{amsmath}
\usepackage{xparse}
\usepackage{braket}
\usepackage{babel}
\usepackage{graphicx}
\usepackage{braket} 
\usepackage{float}
\usepackage{tikz}
\usepackage[T1]{fontenc}
\usepackage{dsfont}

\DeclareMathOperator{\Tr}{\text{Tr}}

\hypersetup{
colorlinks=true,
citecolor=blue,
linkcolor=red,
urlcolor=black
}


\newcommand{\id}{\hat{\mathds{1}}}

\begin{document}

\title{Exact description of the boundary theory of the Kitaev Toric Code with open boundary conditions}

\author{Yevheniia Cheipesh}
\affiliation{
Institut f\"ur Theoretische Physik,
 Georg-August-Universit\"at G\"ottingen, 37077 G\"ottingen, Germany
}

\author{Lorenzo Cevolani}
\affiliation{
 Institut f\"ur Theoretische Physik,
 Georg-August-Universit\"at G\"ottingen, 37077 G\"ottingen, Germany
}

\author{Stefan Kehrein}
\affiliation{
Institut f\"ur Theoretische Physik,
 Georg-August-Universit\"at G\"ottingen, 37077 G\"ottingen, Germany
}

\date{\today}

\begin{abstract}
In this work we consider the Kitaev Toric Code with specific open boundary conditions. Such a physical system has a highly degenerate ground state determined by the degrees of freedom localised at the boundaries. We can write down an explicit expression for the ground state of this model. Based on this, the entanglement properties of the model are studied for two types of bipartition: one, where the subsystem $A$ is completely contained in $B$; and the second, where the boundary of the system is shared between $A$ and $B$. In the former configuration, the entanglement entropy is the same as for the periodic boundary condition case, which means that the bulk is completely decoupled from the boundary on distances larger than the correlation length. In the latter, deviations from the torus configuration appear due to the edge states and lead to an increase of the entropy. 
We then determine an effective theory for the boundary of the system. In the case where we apply a small magnetic field as a perturbation the degrees of freedom on the boundary acquire a dispersion relation. The system can there be described by a Hamiltonian of the Ising type with a generic spin-exchange term.
\end{abstract}

\maketitle

\section{Introduction}

The Landau theory of phase transitions is one of the most precise paradigms in modern physics. Once a system undergoes a symmetry breaking, the macroscopic properties of its ground state change.  Some phase transitions are not characterised by the local order parameter, instead the topological order can be used~\cite{Haldane1983a, Haldane1988, Haldane1983b, Kosterlitz1974, Kosterlitz1972, Kosterlitz1973, Laughlin1981}. The latter plays a crucial role in the existence of so--called topological insulators. In these systems, conduction can take place only along the boundary, while in the bulk one has an insulating state. Topological properties are extremely robust, they cannot be destroyed by simple perturbations. %The predictions obtained using  these arguments are extremely general. %
Many of these states, e.g. the Fractional Hall Effect, involve the correct description of degrees of freedom that are not located in the bulk but on the boundaries of our system. If we consider for example a bi--dimensional system with open boundary conditions, a large number of degenerate ground states will be located on the boundary. These states are called \textit{edge states}~\cite{Kane2005a, Kane2005b, Bernevig2006, Murakami2006, Hasan2010} and their correct description can be complicated. Physically, these states are responsible for the behavior of the system in the low energy regime. The bulk theory is usually gapped, while the boundary theory is known to be gapless, meaning that in the low energy limit the boundary states will affect the behavior of the observables. A remarkable example of such effect is the current quantization in Hall systems~\cite{Laughlin1983, Chang2013, Tsui1999, Zhang2005}.

In order to grasp the physics of such states, exactly solvable models as~\cite{Wen2003a,Wen2003b, Kitaev2006, Kitaev2003} play an important role. These can be used as playground to develop exact mathematical concepts that can then be extended to other more complicated systems. One of these models is the so--called Kitaev Toric Code (KTC)~\cite{Kitaev2003}, whose peculiarity is due to the toric geometry. This model can be used to store quantum information in its 4-fold degenerate ground states with different topological properties~\cite{Kitaev2003, bravyi1998quantum, Corcoles2015}.\\

In this paper we study the KTC with particular open boundary conditions: the coupling constants of some operators are set to zero along the boundary. The model is exactly solvable in this case.  %We demonstrate that the ground state of this model can be written applying boundary and bulk operators to a product state.%
The exact representation of the ground state gives a clear expression of the edge states, which are localized on the boundary of the system, but also traverse the bulk. The specific form of the ground states can be constructed from different operators acting separately on these regions. The resulting state exhibits a clear connection between the bulk and the boundary. We then use the derived exact expression for the ground state to compute the bipartite entanglement entropy of our model. We find that this has the same value as for the periodic boundary condition case if the partition is in the bulk of the system. If the boundary is also partitioned, then the value of the entanglement entropy is increased by a factor that depends on the length of the boundary of a subsystem. Due to the fact that the geometry is topologically trivial, the ground states cannot be used to store information safely as in the original model. In the end, we study how the boundary degrees of freedom can be excited using a magnetic field to induce a non-flat dispersion relation.

The paper is structured as follows: in the Section I we describe the model and the boundary conditions we choose, in the Section II we focus on the construction of the ground state and on its interpretation on the bulk--to--boundary correspondence, the third Section is dedicated to the computation of the entanglement entropy, and in the last Section we focus on the excitations of the boundary theory and their dispersion relation.

\section{The Kitaev Toric code with open boundary conditions}

We want to study a two--dimensional spin system with open boundary conditions (OBC). The system is a $m\times n$ rectangle with spin-$1/2$ located at the links between the lattice sites which Hamiltonian reads as:
\begin{align}\label{eq:Hamiltonian}
\nonumber \hat{H} &= -J_e\sum_v\hat{A}_s - J_m\sum_p\hat{B}_p, \\
\hat{A}_s &= \prod_{\text{star}}\hat{\sigma}^x_i, \qquad \hat{B}_p = \prod_{\text{plaquette}}\hat{\sigma}^z_i.
\end{align}
$\hat{B}_p$ is called "plaquette" operator, and $\hat{A}_s$ is the "star" operator. These two operators trivially commute and can have eigenvalues $\pm 1$. The ground state of this model is an eigenstate for all the star and plaquette operators with eigenvalues $+1$.  The Hamiltonian~\eqref{eq:Hamiltonian} with periodic boundary conditions (PBC) is called the Kitaev Toric Code (KTC) and has already been studied in several works in the literature~\cite{Kitaev2003, Kitaev2006}. Its ground state is 4-fold degenerate and has non--trivial topological properties living on a torus due to the PBC. This model has been proposed to be used as an error--free memory~\cite{bravyi1998quantum, Kitaev2003, Kitaev2006}. Various more complicated models with the same properties have since been discussed in the literature~\cite{Beigi2011, Levin2013}.\\

\begin{figure}[h!]
\includegraphics[scale=0.7]{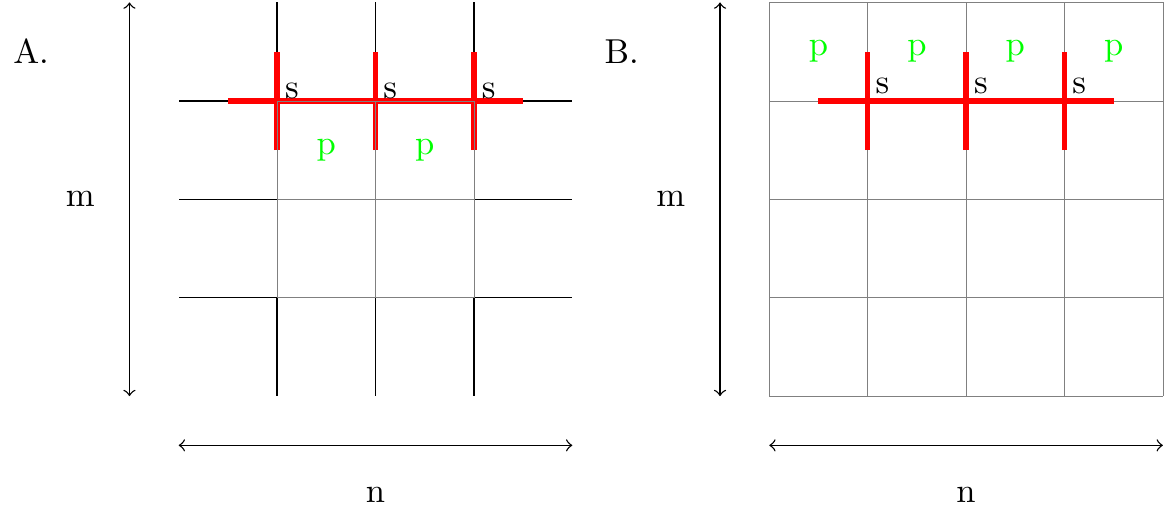}
\caption{In this figure we present the boundary conditions we are going to use. In Panel (A) only star operators act on the boundary of the system, while in Panel (B)  spins on the boundary are constrained only by plaquette operators. The latter configuration, (B), is going to be studied in our work. Despite the fact that these two look quite different, they are connected by a gauge transformation.}
\label{fig:types_of_bc}
\end{figure}

The specific case of the Hamiltonian~\eqref{eq:Hamiltonian} with different OBC has been studied in several cases~\cite{bravyi1998quantum, Yu2013}. These models are referred to
as planar codes. One of the main results of these works is the determination of the edge states which are all degenerate.\\
In our work we are going to consider a modified KTC with particular OBC: we set to zero the $J_e$ constants along the boundary of our system, see Fig.~\ref{fig:types_of_bc} (B) for a graphical representation. This means that the plaquette operators are still active on the boundary, while the star operators are not. We will discuss the physical implications of turning on and off these operators in the next section. Clearly, it is also possible to set the plaquette operators to zero instead, see Fig.~\ref{fig:types_of_bc} (A). From now on we will always refer to the configuration (B), but our results can be extended to the other case by using the eigenvectors of $\sigma_x$ as a local basis instead of ones of $\sigma_z$ as we will do henceforth.\\
%{\color{red}
%Our choice of boundary conditions can be motivated by recent results on entanglement entropy calculations using flow equation technique, see Ref.~\cite{Wegner2000, Kehrein2006}. There, the entanglement entropy can be computed in the limit where the couplings along a line are sufficiently small where the two subsystems are almost decoupled~\cite{Kehrein2017}. The two subsystems are then almost decoupled and a perturbative expression for the mean entropy can be written. In the specific case of the kitaev toric code this technique is not applicable because of the discontinuity of the entanglement entropy at exactly zero coupling constants. Anyways, the physics arising from these models is worth to be studied.
%}
 
% 
%  The ground state in the original KTC in the state with an eigenvalues $+1$ for all operators in \eqref{eq:Hamiltonian}. In other words, each spin is restricted by 2 plaquette and 2 star conditions. Its is known to be 4--fold degenerate. In our case spins on the boundary are restricted only by 1 condition (directly on the boundary) or 3 conditions (spins right next to the boundary). This allows us to suggest the existence of non--trivial highly degenerate ground state.

% Before to start analyzing the  Consider KTC that consists of an $N\times N$ {\color{red} Since the figures are $m\times n$ it would be better to use that notation, or change the figures... one of the two.}  lattice with periodic boundary conditions.  We have  $2N^2$ qubits and $2N^2$ plaquette and star conditions. However, due to the periodicity, $\prod_{s} A_s = \prod_{s} B_s = +1$. So only $2N^2 - 2$ operators are independent and the dimensions of the ground state is $2^2 = 4$.

\section{Ground state properties of the KTC with open boundary conditions}
\begin{figure}[t]
\includegraphics[scale=0.75]{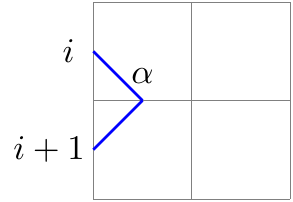}
\caption{Representation of the operator $\mathcal{W}_i$ presented in Eq.~\eqref{eq:W} as a small path connecting two nearest--neighbour sites. We have a freedom of choice for the orientation of the boundary, which is a one--dimensional system with periodic boundary conditions, so it can be mapped into a circle. }
\label{fig:building} 
\end{figure}

 Let us first take a look at the ground state of the original KTC~\cite{Kitaev2003}. The full set of degenerate ground states can be constructed starting from the trivial state when all the spins are down

\begin{equation}
\ket{0}=\prod_{j} \ket{\downarrow}_j
\end{equation}
where $j$ spans the full lattice.\\
Then the ground state of the KTC in presence of PBC has been demonstrated to be \cite{Kitaev2003}:
\begin{equation}
\ket{GS}_a = \sqrt{2}\prod_{j} \left( \frac{\id + \hat{A}_j}{\sqrt{2}} \right)\hat{Q}_a\ket{0}
\end{equation}
where the index $j$ runs over the different lattice sites. The operators $\hat{A}_j$ were defined in Eq.~\eqref{eq:Hamiltonian}. The operator $\hat{Q}_a$ belongs to a set of 4 different operators $\{\id, \hat{S}_1, \hat{S}_2, \hat{S}_1 \hat{S}_2\ \}$, where $\hat{S}_1, \hat{S}_2$ correspond to closed loops that flip spins along two main circles of the torus, correspondingly. One sees that the ground state with PBC is four--fold degenerate.\\
In the case of OBC, the previous state is still a valid ansatz for the ground state. However, it is just one among many degenerate states that is also valid in the case with PBC. The degrees of freedom obtained by removing the constraints on the boundary enrich the ground state subspace  compared to the one  in the original model. \\
%They also belong to the ground state subspace and this can be intuitively understood because with open boundary conditions we allow much more freedom to the spins on the boundary, which means that we expect more configurations with the same energy to appear.
Edge states arise on the boundary creating a huge degeneracy and can be written as:
\begin{equation}\label{eq:belgroundstate}
\ket{GS} = \prod_{j}^{\Sigma}  \left(\frac{ \id+ \hat{A}_j}{\sqrt{2}}\right)\sum_{\vec{e}} c_{\vec{e}}  \prod_{i}^{\mathcal{L}-1} \left( \mathcal{W}_i \right)^{e_i}\ket{0}.
\end{equation}
The operators $\mathcal{W}_i= \sigma_i^x\sigma_{i+1}^x\sigma_\alpha^x$ are represented in Fig.~\ref{fig:building} and they involve just the degrees of freedom located at the boundary. 
The vector $\vec{e}$ has length equal to the boundary length $\mathcal{L}-1$ and every entry can be either $0$, which means no operator $\mathcal{W}_i$ on that point,  or $1$, which means that $\mathcal{W}_i$ is "activated" on that point. One should note, however, that here we consider rectangular lattice which means that we have additionally 4 corners that have a different configuration compared to the sites along the boundary. That is why, strictly speaking, we should also add 4 additional operators $\mathcal{W}_j = \sigma_j^x\sigma_{j+1}^x$ that  have the same nature and structure as ones introduces for the boundary and represented in Fig.~\ref{fig:building}, but involve only 2 spins that are  flipped on the corner. This is necessary in order to full fill the condition that the product of all  $\mathcal{W}_i$ is trivial. In the case of circular geometry one does not have this issue. The number of ground states is then equivalent to the number of different vectors present: $2^{\mathcal{L}-1}$. The coefficients $c_{\vec{e}}$ take into account all the possible superpositions of these boundary states. It is important to notice how the state, previously written in Eq.~\eqref{eq:belgroundstate}, takes the form of an expansion of states that are factorised into a part defined on the boundary, the one involving $\mathcal{W}_i$ operators, and a part defined on the bulk, the one involving $\hat{A}_i$ operators.\\
Eq.~\eqref{eq:belgroundstate} is one of our main results, in the following two sections we will describe how to obtain this state starting from the operators defining Hamiltonian~\eqref{eq:Hamiltonian} in full details. The reader not interested in these details can skip this part and go directly to the Section IV dedicated to the entanglement entropy.

\subsection{Construction of the ground state}

\begin{figure}[h]
\centering
\includegraphics[scale=0.65]{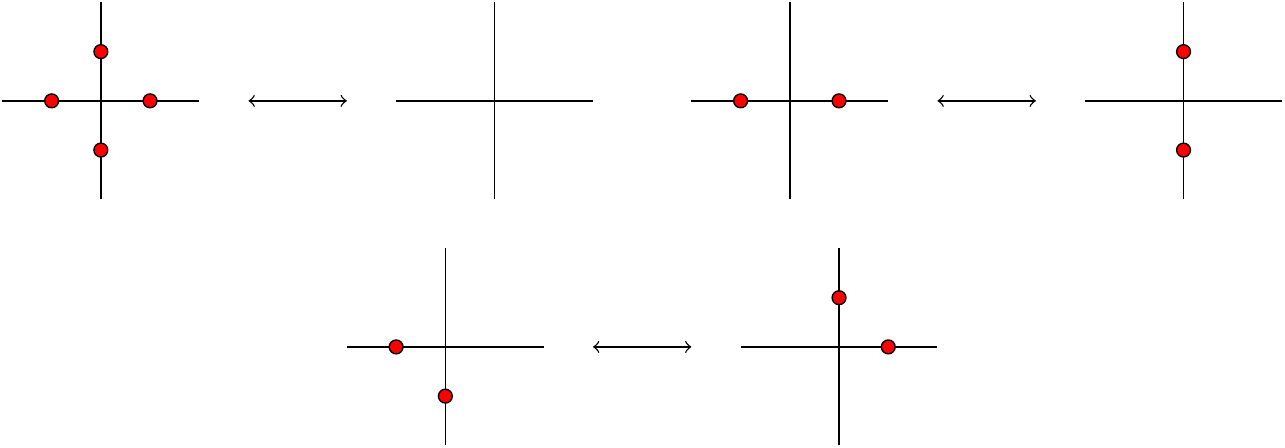}
\caption{In this Figure, we present how the star operator $\hat{A}_s$ acts on some spin configurations. One can see how it is possible to come back to the initial state applying the same star operator twice.}
\label{fig:star}
\end{figure}

Before we focus on the structure of the ground state, we want to demonstrate that it is massively degenerate. The degeneracy of the ground state can be determined by counting the degrees of freedom of the system and then subtracting the number of constraints imposed by the stabilizers~\cite{Legget2013}.
For the sake of simplicity, let us consider an  $N\times N$ lattice. There, the number of q-bits is $2N^2 + 2N$, where  $2N$ comes from the boundary. The number of plaquette operator is $N^2$ while the one of the stars is $N^2 - 2N + 1$, due to the fact that they are absent on the boundary. So the total number of all configurations allowed by the constraints of the ground state is $2^{4N-1}=2^{\mathcal{L}-1}$ where $\mathcal{L}=4N$ is the length of the boundary.\\
Now we want to construct the ground state itself following the requirement that it should be an eigenvector of all $\hat{A}_s, \hat{B}_p$. Such a vector exists as far as $\left[ \hat{A}_s,\hat{B}_p\right] = 0$. We use the local basis of $\sigma_z$, where
\begin{align}\label{statesz}
\ket{\phi_1} = \left( \begin{array}{c} 0 \\ 1 \end{array}\right) \qquad  \ket{\phi_2} = \left(\begin{array}{c} 1 \\ 0 \end{array} \right)
\end{align}

\begin{align}\label{actionz}
\nonumber \hat{\sigma}_x\ket{\phi_1}&=\ket{\phi_{-1}} \qquad \hat{\sigma}_x\ket{\phi_{-1}}=\ket{\phi_{1}} \\ 
\hat{\sigma}_z\ket{\phi_1}&=\ket{\phi_{1}} \qquad \hat{\sigma}_z\ket{\phi_{-1}}=-\ket{\phi_{-1}}.
\end{align}
This is the most suitable choice of local basis for our boundary conditions.\\
Let us now proceed to build the ground state. To do so, we need to have a clear picture of what the star and plaquette operators physically represent.
\begin{figure}[h!]
\centering
\includegraphics[scale=0.75]{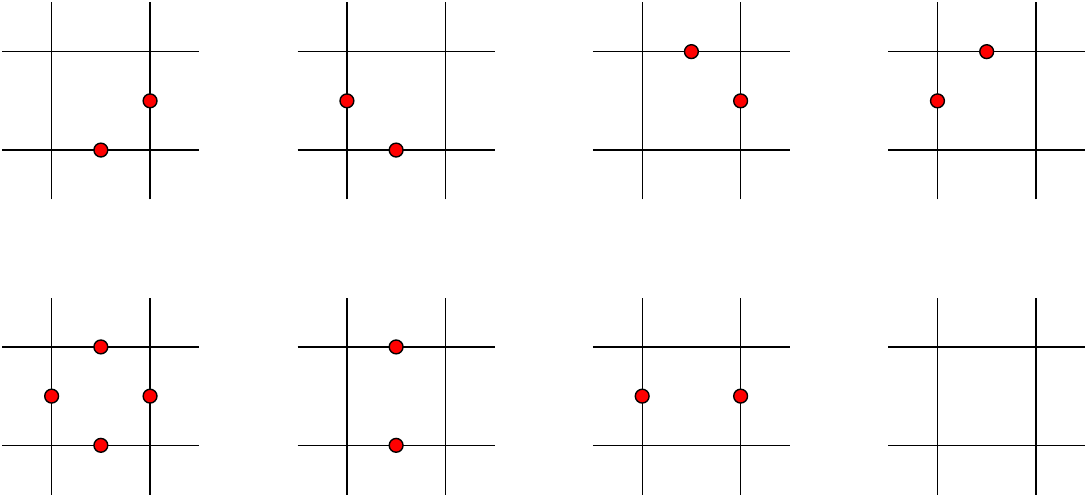}
\caption{In this Figure, we present all the plaquette configurations present in the ground state. The red dots represent spins up, while the plain line represents a spin down. As long as we always have an even amount of spins up, the eigenvalue of the plaquette operator is always positive.}
\label{fig:plaquette}
\end{figure}
It is useful to start from considering $B_p$, since it is diagonal in our local basis. For the unit cell with four spins as in Fig.~\ref{fig:plaquette}, the product of the spins along the square has to be equal to $+1$. Clearly, only the configurations with an \textit{even} number of spins up satisfy the ground state condition. This requirement  will be important also for the construction on the boundary since the plaquette operators are the only ones active there. More interesting is the action of the star operators. Since each star operator is composed of four $\sigma_x$ matrices, it corresponds to a simultaneous flip of all the corresponding spins. One can see the action of star operators on all possible states in Fig.~\ref{fig:star}. In other words, states are transformed into each other under the action of the start operator. It is important to note, that the new state is still an eigenstate of a plaquette operator with an eigenvalue $+1$ since the star operator cannot change the parity of spins.\\

\begin{figure}[h]
\centering
\includegraphics[scale=0.85]{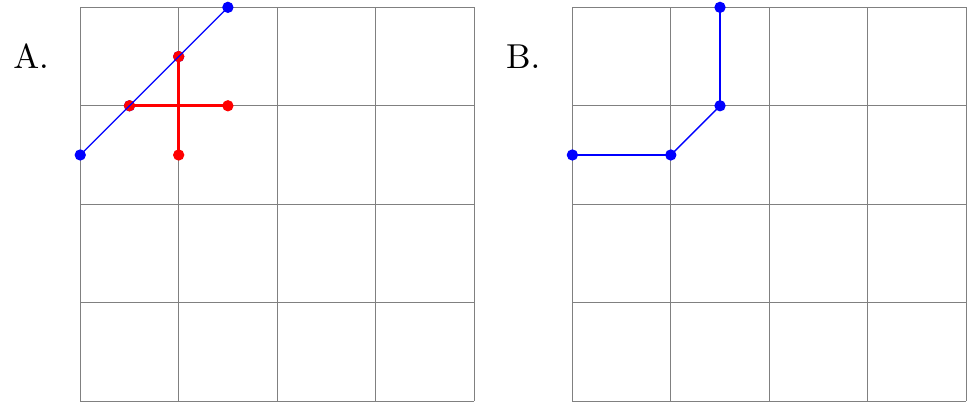}
\caption{In this Figure, we want to present a typical boundary state of our model, where we use the boundary conditions of Fig.~\ref{fig:types_of_bc} (B). One can see how a spin-flip along a line connecting the two boundaries creates a vertex-free state where all the plaquettes have eigenvalue $+1$. In the bulk, the star operators act modifying the path connecting the two points, see for example Panel (B), but they do not modify the points on the boundary.}
\label{fig:gs_vector}
\end{figure}

The previous considerations can be used to construct the requirements for the ground state of the KTC (namely that each plaquette can have only an even amount of spins flipped) and also to understand the action of the star operators on such states.\\
Now we explicitly construct the ground state of the model with OBC. Let us start considering the boundary of our system where only plaquette operators act. Starting from a configuration where all the spins are in the eigenstate $+1$, we can see that the path of spin flips connecting two points on the boundary does not cost any energy as we flip two spins for every plaquette, see Fig.~\ref{fig:gs_vector} (A).  It means that the ground state consists not only of the state $\ket{0}$ where all the spins look up but can also have "paths" of flipped spins that end on the boundary.\\

\begin{figure}[h]
\includegraphics[scale=0.56]{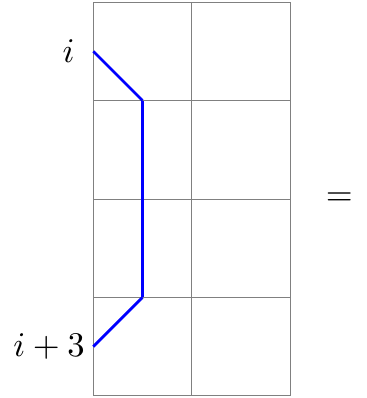}
\includegraphics[scale=0.56]{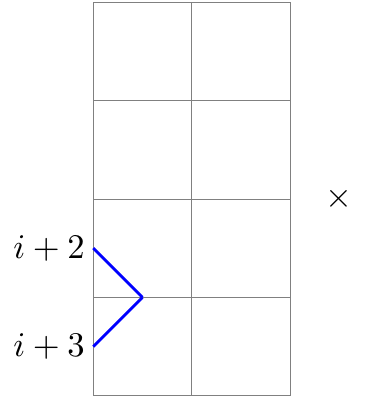}
\includegraphics[scale=0.56]{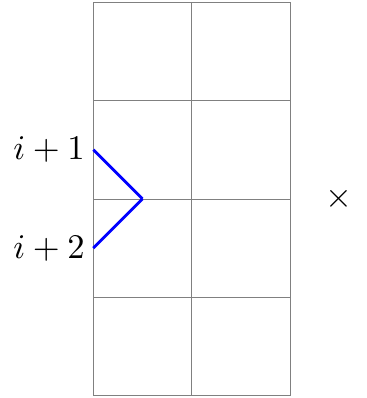}
\includegraphics[scale=0.56]{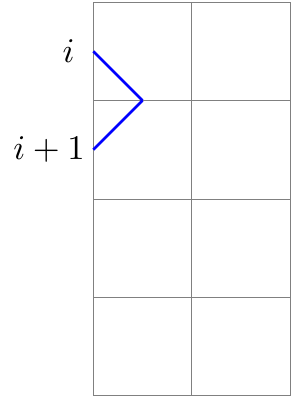}
\caption{\label{fig:building_blocks}Here we give a graphical representation of Eq.~\eqref{eq:Wcomposition}. The $\mathcal{W}_i$ operators can be multiplied to create all the possible configurations on the boundary system.}
\end{figure}

Since the star operators transform states into each other as in Fig.~\ref{fig:star}, the ground state should include all the possible states, allowed by plaquette conditions, with equal coefficients. In this case we stay in the same orbit of the group, generated by all the possible products of the star operators. As far as this group acts only in the bulk of the system, because of our particular boundary conditions, it does not change the spin configuration on the boundary which means that the element of the basis $\ket{l}$ from which the ground state is composed is defined by the particular set of points on the boundary. Thus we can build the ground state as:
\begin{equation}
\label{eq:groundstate}
\ket{GS}=\sum_l c_l \prod_i \left( \hat{\mathds{1}} + \hat{A}_i \right) \ket{l}.
\end{equation}
An important physical interpretation of the star operators can be seen by comparing Fig.~\ref{fig:gs_vector} (A) and (B): applying the star operator to the part of the path that connects the two boundaries changes the state, but \textit{just in the bulk}. This connection between the $\hat{A}_i$ operators and the modification of the paths (\textit{stretching}) can be used to give a different form of Eq.~\eqref{eq:groundstate}.\\
The structure of the deformations of the paths induced by $\hat{A}_i$ can be recast in a more formal way~\cite{Hamma2005}. We introduce the group $\mathcal{A}$ that contains all the possible stretches of any given path $\ket{l}$. The elements of $\mathcal{A}$ are operators $\hat{g}$ generated by all possible products of the star operators in the bulk of the system. The ground state can then be rewritten as a superposition of the action of all the elements of this group:
\begin{equation}\label{eq:groundstate_with_group}
\ket{GS} = \frac{1}{\sqrt{\vert \mathcal{A} \vert}} \sum_l a_l \sum_{\mathcal{A}} \hat{g}\ket{l},
\end{equation}
where we used the fact that any configuration of the star operators is contained in $\mathcal{A}$, meaning that $\hat{g}\in \mathcal{A} \rightarrow \hat{A}_i \hat{g} = \hat{g}^\prime$, and we normalised the vector using $\vert \mathcal{A}\vert=2^{\Sigma}$ where $\Sigma$ is the total number of independent star operators. Eq.~\eqref{eq:groundstate_with_group} is even easier to interpret physically than Eq.~\eqref{eq:groundstate}: the boundary and the bulk are completely decoupled due to the absence of stars that can stretch spin flips from the bulk to the boundary and vice versa.
\\
Eq.~\eqref{eq:groundstate_with_group} says that if we have two points on the boundary, we can connect them and, in order to be a ground state, the resulting state should be a superposition of all the possible "stretches" of the paths between these two points. We can identify them as one equivalence class, namely, we say that two configurations are equivalent if they correspond to the same set of points on the boundary. This allows us to work only with one element of each class and detaches ourselves from the consideration of the bulk.\\
This particular set of paths is represented in Fig.~\ref{fig:representor}, where we simply connect the two points on the bulk by the path closest to the boundary. The action of the elements of $\mathcal{A}$ will take care of modifications in the bulk to have a proper ground state. \\
We want to focus now on the boundary states $\ket{l}$. We will demonstrate that it is possible to create all of them starting from the simple building blocks. One of these building blocks is shown in Fig.~\ref{fig:building} which is represented by three spin flips connecting two nearest--neighbour spins on the boundary:
\begin{figure}[h!]
\centering
\includegraphics[scale=0.7]{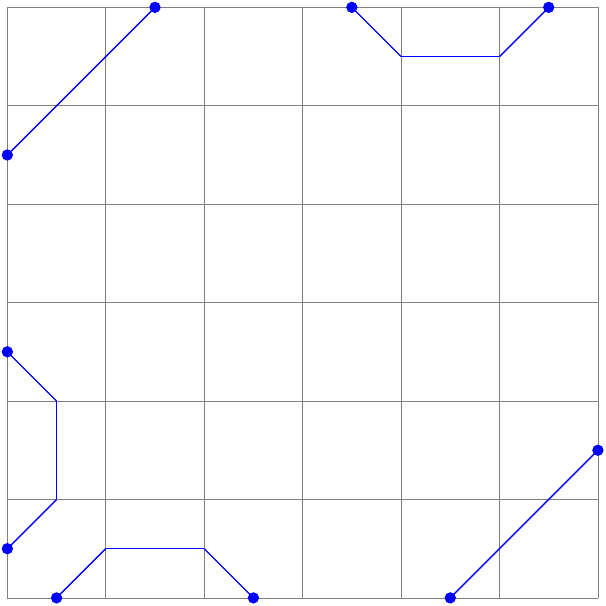}
\caption{A generic boundary state which forms a particular basis element of the ground state. Because of the action of star operators, the path connecting the boundary points is not important, so we can choose by convention the one closest to the boundary. Even if this state looks extremely complicated, this and all the others can be constructed starting from the building block presented in Eq.~\eqref{eq:W}.}
\label{fig:representor}
\end{figure}
%\begin{figure}[h]
% \scalebox{0.7}{\begin{tikzpicture}
%\draw[help lines] (0,0) grid (2,2);
%\draw [blue, thick]  (0,0.5) -- (0.5,1);
%\draw [blue, thick]  (0.5,1) -- (0,1.5);
%\node at (-0.25,1.5) {$i$};
%\node at (.5,1.25) {$\alpha$};
%\node at (-0.45,0.5) {$i+1$};
%\end{tikzpicture}}
%\caption{\label{fig:building} Representation of the operator $\mathcal{W}_i$ presented in Eq.~\eqref{eq:W} as small path connecting two nearest-neighbour sites. The freedom of choice we have in its definition is the orientation of the boundary, which is a one dimensional system with periodic boundary conditions, so it can be mapped into a circle. }
%\end{figure}
\begin{equation}\label{eq:W}
\mathcal{W}_{i} = \sigma_i^x \sigma_{i+1}^x \sigma_\alpha^x,
\end{equation}
where $i$ denotes the spins on the boundary and $\alpha$ the only one in the bulk.\\
The boundary of a two--dimensional system can be represented exactly by a one--dimensional system with periodic boundary conditions. All the physical quantities will depend just on the distance between $i$ and $j$, which label the points on the boundary, and never on $i$ and $j$ alone.
These operators can be used to construct any type of boundary theory simply by multiplying them appropriately	. Let us take the example of Fig.~\ref{fig:building_blocks} where a segment connecting two points is created:
\begin{equation}\label{eq:Wcomposition}
\mathcal{W}_{i}\left( R=3 \right) = \mathcal{W}_{i} \mathcal{W}_{i+1} \mathcal{W}_{i+2}.
\end{equation}

We can check that all the ground states are built by these operators. If the boundary length is $\mathcal{L}$, we have $\mathcal{L}$ $\mathcal{W}_i$ operators. Each of them can either be activated or not, creating different types of boundary states. The total number of possibilities is then $2^{\mathcal{L}}$. Also, if we look at the boundary alone, the state where all the $\mathcal{W}_i$ are applied is the same to the identity, because:
\begin{equation}
\prod^{\mathcal{L}}_i \mathcal{W}_{i}= \prod^{\mathcal{L}}_i \sigma_i^x\sigma_{i+1}^x\sigma_\alpha^x= \hat{\mathds{1}}_{\text{boundary}}\left(\prod_{\text{Bulk}} \sigma_\alpha^x\right).
\end{equation}
This reduces the number of possible boundary states from $2^{\mathcal{L}}$ to $2^{\mathcal{L}-1}$, which coincides with the number of ground states we obtained computing the constraints and degrees of freedom.\\

All the boundary states can then be written as functions of numbers $e_i$ that are $0$ if the corresponding $\mathcal{W}_i$ is not applied to $\ket{0}$ and $1$ if it is applied. This numbers can then be stored into a vector $\vec{e}$ which then represents the boundary theory as:
\begin{equation}
\ket{l} \rightarrow \vec{e} \quad \ket{l}= \prod_{i}^{\mathcal{L}} \left( \mathcal{W}_i \right)^{e_i} \ket{0}.
\end{equation}
As we said before, the boundary conditions can restrict the dimension of these spaces. The boundary, when considered as a system by itself, has a pure periodic boundary conditions, meaning that the vector where all the $\mathcal{W}_i$ are present is the same to the one where none of the spins is flipped.
\begin{equation}
\left(0,0,0, \dots,0,0\right) \sim \left( 1,1,1\dots,1,1 \right),
\end{equation}
which means that the effective number of degrees of freedom is $\mathcal{L}-1$ and the total dimension of the space is then $2^{\mathcal{L}-1}$ which is exactly the total number of ground states of the model.\\

We can now focus on the coefficients $a_l$ of the expansion in Eq.~\eqref{eq:groundstate_with_group}. Since the boundary of our system is topologically equivalent to a one--dimensional system with PBC, a lot of these coefficients will be equal due to the translational symmetry. We can start from the fundamental building blocks, Eq.~\eqref{eq:W} and Fig.~\ref{fig:building}. Since all these operators are equal, their coefficients have to be equal too: $a_1=a$.
The fact that all the other states in the superposition can be obtained by the multiplication of these fundamental blocks fixes the other coefficients. Since the multiplication of operators $\mathcal{W}_i$  gives more complicated boundary state, the generic element of the expansion is given by
\begin{equation}
\ket{l}= \prod_{i}^{\mathcal{L}} \left( \mathcal{W}_i \right)^{e_i} \ket{0} \quad \rightarrow \quad a_l = a^l e^{i\varphi_l},
\end{equation}
where $l=\sum_i e_i^2=\vert \vec{e}\vert^2$, $\varphi_l$ is the phase of the complex number, we fixed $\varphi_1=0$. Physically, this means that the states with the same number of operator $\mathcal{W}_i$ differ just by a phase.\\

We want to emphasise here, that even though the considerations made in this section were formulated for a square geometry with no star-operator on its boundary. Our conclusions can be easily extended to more exotic geometries and boundary conditions where some of the operators of the Hamiltonian are set to zero along some parts of the boundary. The ground state can be constructed starting from the $\mathcal{W}_i$ operators applied over a part of the boundary and by the application of the bulk operators. The dimension of the ground state depends on the \textit{length} of the boundary where one has removed the stabiliser operators.

\section{Entanglement entropy of the ground state}
\label{sec:EEGS}

Once we have the exact analytical expression for the ground state, we can now compute  the entanglement entropy of a bipartition of the system. The system is in a pure state, of which the density matrix is
\begin{equation}
\hat{\rho}=\ket{GS}\bra{GS}.
\end{equation}
At zero temperature, the entanglement in a many--body system can be estimated using the von Neumann entropy~\cite{Amico2008, Horodecki2009}:
\begin{equation}
S_A = -\Tr_A \hat{\rho}_A \ln \hat{\rho}_A,
\end{equation}
where $\hat{\rho}_A=\Tr_B\hat{\rho}$. The entanglement entropy of the KTC has been computed exactly in \cite{Hamma2005, Hamma2005b}
\begin{equation}
S_A \equiv \tilde{S}_A = \left(\mathcal{L}_{A/B}-1\right)\ln 2.
\end{equation}
These calculations show that the entanglement entropy follows the area law, as expected from the presence of a finite correlation length~\cite{Srednicki1993, Eisert2010}, plus a constant term which can be computed exactly. This constant term is the topological entanglement entropy~\cite{Kitaev2006topo, Levin2006,shi2018characterizing}, and it is connected to the quantum dimensions of the anyons present in the theory.\\
We want to compute the ground state entanglement entropy for our system with the specific boundary conditions specified in Fig.~\ref{fig:types_of_bc}. The rectangular lattice is split into two sub--systems (A and B) as presented on the left panel of Fig.~\ref{fig:partition} so that every spin lies either in sub--system A or B, no spins are shared between them. With respect to the KTC, we have more freedom in the choice of the partition between the subsystems $A$ and $B$. In the particular case when a subsystem $A$ is smaller than the subsystem $B$, we can consider two main configurations: the first is where the boundary of the system is completely contained in the subsystem $B$, and the second is when the two subsystems share part of the boundary. \\
The first case is similar to the case of the KTC with PBC. There, the information about the boundary states is completely destroyed by the trace over the subsystem $B$ (we are assuming that $A$ is at least one correlation length away from the boundary). This and the fact that the bulk and the boundary are completely decoupled due to our boundary conditions makes it intuitive that the entanglement entropy takes the same form as for PBC, $S_A=\tilde{S}_A$.\\
However, when  also the boundary is partitioned, we have an increase in the entanglement entropy due to the edge states:
\begin{equation}
S_A = \tilde{S}_A + \ln \dfrac{1}{ f(A)f(B)},
\end{equation}
where the function $f(x)$ is a positive monotonously increasing function that depends only on the length of the boundary in the subsystem.\\
These are the main results of the section.The reader interested in the detailed derivation is encouraged to follow the next subsection. In other case, proceed to section V.
\subsection{Entanglement entropy of the boundary}
\begin{figure}[h!]
\includegraphics[scale=0.95]{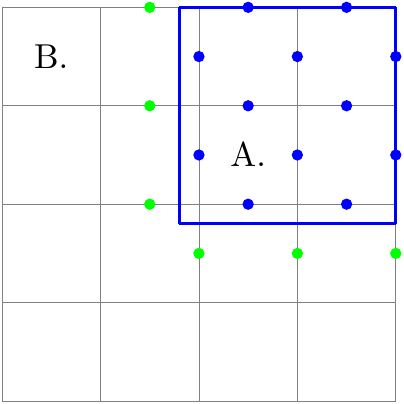}
\quad 
\includegraphics[scale=0.64]{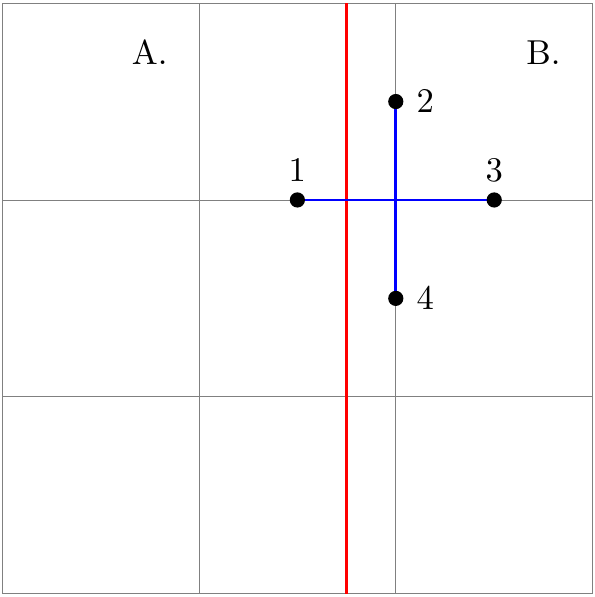}
\caption{\label{fig:partition}Left Panel: Partition of the full system into two subsystems $A$ and $B$ sharing the boundary of the full system with OBC. Right Panel: Degrees of freedom of a star, in this case, shared between the $A$ and $B$ subsystems.}
\end{figure}

We now want to focus on the second case, where two subsystems share part of the boundary, see Fig.~\ref{fig:partition}. In this case, the boundary states will not be completely traced out performing $\Tr_B$, but they will leave their fingerprint on the entanglement entropy.

We start representing the ground state using the group $\mathcal{A}$:

\begin{equation}
\ket{GS} = \frac{1}{\sqrt{\vert \mathcal{A} \vert}} \sum_{l}a^{n_l}e^{i\varphi_l}\sum_{g \in \mathcal{A}}g  \prod_{i}^{\mathcal{L}} \left( \mathcal{W}_i \right)^{e_i} \ket{0}.
\end{equation}

Let's first calculate the reduced density matrix. Before doing this, we should specify carefully what happens to the stars that are shared by two subsystems A and B (see Fig.~\ref{fig:partition}). We decided to place the boundary of the two subsystems in between the spin and the lattice site. In doing this we cut some of the star operators between the two subsystems, see Fig.~\ref{fig:partition}. Once we trace out the degrees of freedom of $B$, this operation fixes part of the spins in $A$, due to the ground state conditions. Keeping track of this reduction is fundamental to take properly into account the degrees of freedom in between $A$ and $B$, which is then responsible of the entanglement entropy.\\
We start by tracing out the degrees of freedom of the subsystem $B$ to obtain the reduced density matrix of the system $A$:
\begin{widetext}
\begin{equation}
\hat{\rho}_A = \frac{1}{\vert \mathcal{A} \vert}\sum_{l,k}a^{n_l}a^{n_k}e^{i(\varphi_l - \varphi_k)}\sum_{g,g' \in \mathcal{A}}\Tr{}_B \left(\bra{0}_B \prod_{i}^{\mathcal{L}} \left( \mathcal{W}_i \right)^{e_i} g_B\overbrace{g_{A/B}}^{\text{shared}}g_A\ket{0}_A \bra{0}_Ag'_A\overbrace{g'_{A/B}}^{\text{shared}}g'_B  \prod_{i}^{\mathcal{L}} \left( \mathcal{W}_i \right)^{f_i}\ket{0}_B\right).
\end{equation}
\end{widetext}
The previous expression can be simplified using two important facts: The boundary states in $B$ subsystem are orthogonal, and the bulk states are orthogonal if their modifications are different, meaning:

\begin{align}
& \bra{0} \prod_{i}^{\mathcal{L}} \left( \mathcal{W}_i \right)^{e_i}  \prod_{i}^{\mathcal{L}} \left( \mathcal{W}_i \right)^{f_i} \ket{0} = \prod_i \delta_{e_i,f_i} \\
& \bra{0} g_B^\prime g_B \ket{0} =\delta\left( g_B^\prime -g_B \right) .
\end{align}
Using these identities the previous equation can be simplified to
\begin{align}
\nonumber & \hat{\rho}_A =
\frac{1}{\vert \mathcal{A} \vert}\sum_{l,k}\sum_{g_B,g_B^\prime \in \mathcal{A}_B}\delta^{n^B_l}_{n^B_k}\delta(g_B-g'_B) a^{n_l}a^{n_k}e^{i(\varphi_l - \varphi_k)} \\
& \sum_{g_A,\,g_A^\prime \in \mathcal{A}_A}g_A\prod^{A/B}\hat{\sigma}^x \prod_{i}^{\mathcal{L_A}} \left( \mathcal{W}_i \right)^{e_i}\ket{0}_A\bra{0}_A \prod_{i}^{\mathcal{L_A}} \left( \mathcal{W}_i \right)^{f_i}\prod^{ A/B}\hat{\sigma}^xg^\prime_A.
\end{align}
Where we used the fact that $\ket{0}=\ket{0}_A\otimes \ket{0}_B$ because it is a product state.\\
We can then divide the full group of modification of the paths in two different subsets $\mathcal{A}_A$ and $\mathcal{A}_B$ containing all the stars \textit{fully} contained in the subsystems $A$ and $B$:
\begin{align}
\nonumber &\mathcal{A}_A \equiv \{ \hat{g} \in \mathcal{A}| \hat{g} =\hat{g}_A\otimes  \hat{\mathds{1}} \} \\
\nonumber &\mathcal{A}_B \equiv \{ \hat{g} \in \mathcal{A}| \hat{g} = \hat{\mathds{1}}\otimes \hat{g}_B \}.
\end{align}
The number of elements in these groups is the total number of possible modifications of the state induced by the stars. We can then define the order of these groups as:
\begin{align}
& \nonumber d=\vert \mathcal{A}\vert = 2^{\Sigma} \quad \Sigma=\text{number of stars in the systems} \\
& \nonumber d_A=\vert \mathcal{A}_A\vert = 2^{\Sigma_A} \quad \Sigma_A=\text{number of stars fully in $A$} \\
& \nonumber d_B=\vert \mathcal{A}_B\vert = 2^{\Sigma_B} \quad \Sigma_B=\text{number of stars fully in $B$}.
\end{align}
The presence of stars not fully contained neither in $A$ nor in $B$ makes $\Sigma \neq \Sigma_A +\Sigma_B$. This can be understood from Fig.~\ref{fig:partition} (right). The star in the figure depends on a spin in $A$ and on three spins in $B$. If we perform a trace on one of the two subsystems, its value will then not be completely fixed. To be more precise, the difference in these two quantities is proportional to the length of the boundary between two sub--systems. For the same reason, the deformation operator $\hat{g}$ can be factorized as:
\begin{equation}
\hat{g}=\hat{g}_A \otimes \hat{g}_{A/B} \otimes \hat{g}_B.
\end{equation}
It is useful to introduce the short-hand notation  
\begin{equation}
G = \sum_{\hat{g} \in \mathcal{A}}\hat{g}\prod_{i}^{A/B}\hat{\sigma}^x_i,
\end{equation}
which relates to the operators acting on the $A$ subsystems outside the boundary. \\
The sum over the possible modifications inside $B$ can be performed trivially and it gives a term proportional to the order of the group
\begin{align}
& \sum_{g_B\, g_B^\prime} \delta\left( g_B-g_B^\prime \right) = \sum_{g_B}\hat{\mathds{1}}= d_B.
\end{align}
The summation over the boundary states, labelled by $l$, is a little bit more involved but, in general, the states with the same boundary configuration over $B$ are traced out:
\begin{equation}
\sum_{l,k} a_l a_k^\ast \delta_{n_l^B}^{n_k^B}= \left( \sum_{l_B}\vert a^{n_{l}^B} \vert^2 \right) \sum_{l_A, k_A} a^{n_l^A} a^{n_k^A}e^{i(\varphi_l^A - \varphi_k^A)}.
\end{equation}
We can then define
\begin{equation}\label{eq:fofB}
\sum_{l_B}\vert a^{n_{l}^B} \vert^2= f(B).
\end{equation}
The function $f(B)$ depends just on the boundary of $B$ and on the links between $A$ and $B$  and it is equal to $1$ when it contains all the boundary as far as $\sum_{l} \vert a^{l_n} \vert^2 = 1$ for the whole system. From Eq.~\eqref{eq:fofB}, it is clear that $f(B)\leq 1$ and it is an increasing function of the argument.\\

For our future calculations it is useful to work out $G\ket{0}\bra{0}G\tilde{G}\ket{0}\bra{0}\tilde{G}$:
\begin{align}
& \bra{0}G\tilde{G}\ket{0} = \bra{0}\sum_{\hat{g} \in \mathcal{A}}\hat{g}\prod_{i}^{A/B}\hat{\sigma}^x_i \sum_{\hat{g} \in \mathcal{A}}\hat{g}'\prod_{j}^{A/B'}\hat{\sigma}^x_j \ket{0} .
\end{align}
In order to have non--zero matrix element, we should have
\begin{equation}
\sum_{\hat{g} \in \mathcal{A}}\hat{g}\prod_{i}^{A/B}\hat{\sigma}^x_i \sum_{\hat{g} \in \mathcal{A}}\hat{g}'\prod_{j}^{A/B'}\hat{\sigma}^x_j = \hat{\mathds{1}}.
\end{equation}
This condition can be satisfied in two ways:
\begin{enumerate}
\item $\prod_{i}^{A/B}\hat{\sigma}^x_i \prod_{j}^{A/B'}\hat{\sigma}^x_j = \hat{\mathds{1}}$. In this case two sets of $\hat{\sigma}^x $ are identical. Then we should also have $\sum_{\hat{g} \in \mathcal{A}}\hat{g} \sum_{\hat{g}' \in \mathcal{A}}\hat{g}' = \hat{\mathds{1}}$ and there are $|d_A|$ such possibilities.
\item $\prod_{i}^{A/B}\hat{\sigma}^x_i \prod_{j}^{A/B'}\hat{\sigma}^x_j  = \hat{S}$, where $\hat{S} = \prod_i^{ A/B} \hat{\sigma}^x_i$ taken along all the boundary between A and B. In this case, we say that two sets of $\hat{\sigma}^x_i $ are complementary and together cover all the boundary. Group elements $\hat{g}$ should then satisfy the next condition:  $\sum_{\hat{g} \in \mathcal{A}}\hat{g} \sum_{\hat{g}' \in \mathcal{A}}\hat{g}' =\hat{S}$. There are also $|d_A|$ such possibilities.
\end{enumerate}
It is clear, that in any other case (when we have some spins excited along the $A/B$ boundary), this is not the element of group $\mathcal{A}$ and therefore can not be equal to $\sum_{\hat{g} \in \mathcal{A}}\hat{g} \sum_{\hat{g} \in \mathcal{A}}\hat{g}'$ which leads to zero matrix element. Now we can see that 
\begin{equation}
G\ket{0}\bra{0}G\tilde{G}\ket{0}\bra{0}\tilde{G} = |d_A|\ket{0}\bra{0}G(\hat{\mathds{1}} + \hat{S}).
\end{equation}
Now our reduced density matrix can be written in a more elegant way:
\begin{align}
&\hat{\rho}_A = \frac{d_B f\left( B \right)}{\mathcal{A}}\times\\
&\times\underbrace{\sum_{l,k}a^{l+k} e^{i(\varphi_l-\varphi_k)}}_{\text{from A}}\hat{G}\prod_{i}^{\mathcal{L_A}} \left( \mathcal{W}_i \right)^{e_i}\ket{0}_A\bra{0}_A\prod_{i}^{\mathcal{L_A}} \left( \mathcal{W}_i \right)^{f_i}\hat{G}.
\end{align}
It is also easily seen that
$
(\hat{\mathds{1}} + \hat{S})^2 = 2(\hat{\mathds{1}} + \hat{S})
$.
The expression of the density matrix alone is not enough to compute the entanglement entropy. In order to compute the logarithm of this operator we need to compute all its integer powers.\\

Starting with the square of $\hat{\rho}_A$ we have:
\begin{align}
\nonumber \hat{\rho}^2_A&= d_A d_B^2\dfrac{f(B)^2}{\vert\mathcal{A}\vert^2}\underbrace{\sum_k a^{2k}}_{=f(A)}\sum_{l,k}a^{l+k} e^{i\varphi_l - i\varphi_k} \\
\nonumber &\hat{G} \left[ \prod_{i}^{\mathcal{L_A}} \left( \mathcal{W}_i \right)^{e_i}\ket{0}_A \bra{0}_A \prod_{i}^{\mathcal{L_A}} \left( \mathcal{W}_i \right)^{f_i}\right]\hat{G}(\hat{\mathds{1}} + \hat{S}) \\
\nonumber &= d_A d_B^2\dfrac{f(B)^2f(A)}{\vert\mathcal{A}\vert^2}\sum_{l,k}a^{l+k} e^{i\varphi_l - i\varphi_k} \\
\nonumber &\hat{G} \left[ \prod_{i}^{\mathcal{L_A}} \left( \mathcal{W}_i \right)^{e_i}\ket{0}_A \bra{0}_A \prod_{i}^{\mathcal{L_A}} \left( \mathcal{W}_i \right)^{f_i} \right]\hat{G}(\hat{\mathds{1}} + \hat{S}) \\
&=f(B)f(A)\dfrac{d_A d_B}{d}\hat{\rho}_A(\hat{\mathds{1}} + \hat{S}).
\end{align}
We can then proceed to higher powers:
\begin{align}
\hat{\rho}^3_A &= f(B)f(A)\dfrac{d_A d_B}{d}\hat{\rho}^2_A(\hat{\mathds{1}} + \hat{S})= \nonumber \\
&= 2 \left( f(B)f(A)\dfrac{d_A d_B}{d} \right)^2\hat{\rho}^2_A.
\end{align}
For the general power of $\hat{\rho}_A$:
\begin{equation}
\hat{\rho}^n_A = \left( 2f(B)f(A)\dfrac{d_A d_B}{d}\right)^{n-2}\hat{\rho}^2_A.
\end{equation}
Expanding the logarithm in power series it is possible to compute the entanglement entropy as
\begin{equation}
 S_A = \ln \left( \frac{d}{2d_A d_B} \frac{1}{f(A)f(B)}\right).
\end{equation}
Let us take a look at the expression for the entanglement entropy when the subsystem A is entirely in the bulk. It can be easily done as far as this is just a particular case of the calculation above where the whole boundary is traced out. This would lead to  $\sum_{l} \vert a^{l_n} \vert^2 = 1$ already on the stage of calculating the first power of $\hat{\rho}_A$ due to the orthogonality. In the end the expression will be the same but with $f(A)f(B) = 1$\
\begin{equation}\label{eq:EEWOB}
\tilde{S}_A = \ln \left( \frac{d}{2 d_A d_B}\right).
\end{equation}
Taking into account that $d = 2^\Sigma, d_A = 2^{\Sigma_A}, d_B = 2^{\Sigma_B}$ and $\Sigma = \Sigma_A + \Sigma_B + \Sigma_{A/B}$, \begin{equation}
\tilde{S}_A= \left(\mathcal{L}_{A/B} - 1\right) \ln 2,
\end{equation}
which is exactly the result for KTC with PBC.
 Thy we have:
\begin{equation}\label{eq:EEWB}
S_A = \tilde{S}_A + \ln \dfrac{1}{ f(A)f(B)}.
\end{equation}
The factor $f(A)f(B)$ is always smaller or equal to  $1$. It depends just on the portions of boundary present in the subsystems $A$ and $B$. This contribution leads to the additional term $-\ln f(A)f(B)$ which \textit{increases} the entropy.
\subsection{Entanglement entropy in the bulk}
We can now briefly describe the specific case of $A$ completely contained inside the bulk of the system. There, the boundary is completely traced out by the first trace procedure, which then ends up in the reduced density matrix:
\begin{equation}
 \rho_A = \frac{d_B}{d}\hat{G}\ket{0}\bra{0} \hat{G},
\end{equation}
which gives us the entanglement entropy
$
 S_A =\tilde{S}_A 
$
that is exactly the value of the entanglement entropy with PBC. As we explained in the previous sections, the Hilbert spaces of the bulk and the boundary are decoupled. If the system $A$ is completely contained in the bulk, the information about the boundary is then destroyed once the trace over the subsystem $B$ is performed. The density matrix of $A$ is then the same as in the case of PBC, together with the entanglement entropy. The result is as expected as the boundary is entangled.\\
As a conclusive remark, it is important to notice a significant property of the previous expressions: Eq.~\eqref{eq:EEWB} and~\eqref{eq:EEWOB} are both completely independent of the value of the couplings. Their values are fixed by the particular geometry we choose for the subsystems $A$ and $B$.

\section{Excitation of the boundary theory}
\begin{figure}[h]
\includegraphics[scale=1]{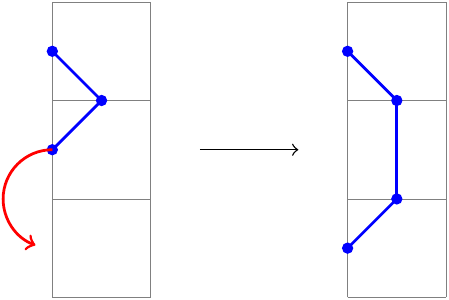}
\caption{Macroscopic tunnelling effect on the boundary theory due to the magnetic field. This perturbation allows the excitations on the boundary to travel with a dispersion relations that can be computed using perturbation theory at the leading order in $h_x/J_m$. For small magnetic field the transitions are just nearest--neighbour because the energy needed is exponential in the number of spin flips. As the intensity of the perturbation is increased, the tunnelling can affect more distant sites, meaning that we have a "non-local" Hamiltonian.
}
\label{fig:transition} 
\end{figure}

Hamiltonian~\eqref{eq:Hamiltonian} has different excitation energies in the bulk and on the boundary due to the different operators acting on these two distinct regions. If we act on the ground state with a generic spin flip, via a $\sigma_i^x$ operator, it will cost an energy equal to $2\left( J_m +J_e \right)$ because every spin is shared between two star and two plaquette operators. On the boundary, we have fewer operators acting on our spins, so if we flip the spin at the boundary we need to pay just $J_m$ because just one plaquette operator is acting on this degree of freedom.\\
Due to the particular structure of the ground state on the boundary, described by $\mathcal{W}_i$ operators, one can see that only two possible excited states are possible: $\sigma_{i}^x \mathcal{W}_i$ and $\sigma_{i}^x \sigma_{i+1}^x\mathcal{W}_i$ due to the spin-$1/2$ algebra. All these excited states are fixed to the points where they are created due to the absence of star operators on the boundary, which allows us to conclude that their dispersion relation is flat.\\
In order to give our boundary theory non-trivial dynamics, we need to have some dispersion relation. This has been done in different topological error correcting codes as the Wen plaquette and the KTC with different OBC by adding a magnetic field~\cite{Yu2013}.
In our case, we also have a magnetic field along x or y--direction which allows us to flip the spins.
\begin{equation}\label{perturbation}
\hat{H}_I = h_x\sum_i \hat{\sigma}^x_i
\end{equation}
Using $\mathcal{H}_I$ as a perturbation of~\eqref{eq:Hamiltonian}, it is possible to compute the tunneling processes induced by the perturbation, see Ref.~\cite{Kou2009, Kou2009b, Yu2009}.
We will now compute the tunnelling between the sites $n=i$ and $m=i+R$ induced by the magnetic field, as represented in Fig.~\ref{fig:transition}

Following \cite{Yu2013}, we can compute the transition $\ket{m} =\sigma_m^x \ket{GS}\rightarrow \ket{n}=\sigma_n^x\ket{GS}$ using the Gell-Mann-Low theorem~\cite{GellMann1951}.\\
The energy shift is defined as
\begin{align}
& \nonumber E = \bra{n}\hat{H}\hat{U}_I(0,\infty)\ket{m} = E_0 + \delta E \\
& \delta E= \bra{n}\hat{H}_I\hat{U}_I(0,\infty)\ket{m} = \bra{n}\hat{H}_I\sum_j\left(\dfrac{\hat{H}_I}{E - \hat{H}_0}\right)^j\ket{m} \label{eq:deltaE}
\end{align}
In order to have a well--defined power series expansion, we will assume $ J_m \gg \vert h_x\vert $, in this way a perturbative expansion of the previous exact expression is motivated. This is particularly important in order to understand the physics behind these processes. When no magnetic field is activated, the part of the $\mathcal{W}_i$ operator that can be modified by the stars, namely the spin--flip is labelled by $\alpha$ in Eq.~\eqref{eq:W} and Fig.~\ref{fig:building}. When the magnetic field is activated, every spin flip costs energy. The paths now have different energies depending on their length, this means that we reduce some degeneracy and the shortest paths connecting two points are energetically convenient.\\
We can start with the nearest--neighbour hopping $\ket{i} \rightarrow \ket{i+1}$ which is clearly $i$--independent because of the translational invariance of the boundary. These states can be represented using the $\mathcal{W}_i \left( R \right)$ operators defined before, since they are also the shortest paths connecting two points on the boundary. We then want to compute the transition probability of the process:
\begin{equation}
\ket{R=0} = \mathcal{W}_i \ket{0}\quad \rightarrow \quad \ket{R=1} = \mathcal{W}_i \left( 1 \right) \ket{0}.
\end{equation}
We can compute the energy shift at the leading order as:
\begin{equation}
\delta E = \bra{R=1} \hat{H}_I \left( \frac{\hat{H}_I}{E-\hat{H}_0} \right)^2 \ket{R=0} = \frac{h_x^3}{\left( -2J_m \right)^2}.
\end{equation}
An easy way to compute these transitions is using a graphical approach, we draw the lines corresponding to the generic states $\ket{R}$ and for every application of $\hat{H}_I$ we flip one spin
\begin{align}
\delta E &= \bra{0}\mathcal{W}_i\mathcal{W}_{i+1} \hat{H}_I\left(\dfrac{\hat{H}_I}{E - \hat{H}_0}\right)^2\mathcal{W}_i\ket{0}\\
& = \dfrac{h_x}{(-2J_m)} \bra{0}\mathcal{W}_i\mathcal{W}_{i+1} \hat{H}_I\left(\dfrac{\hat{H}_I}{E - \hat{H}_0}\right)\hat{\sigma}_x^{\alpha}\hat{\sigma}_x^{i}\ket{0} \\
& = \dfrac{(h_x)^2}{(-2J_m)^2} \bra{0}\mathcal{W}_i\mathcal{W}_{i+1} \hat{H}_I\hat{\sigma}_x^{\beta}\hat{\sigma}_x^{i}\mathcal{W}_i\ket{0}= \dfrac{(h_x)^3}{(-2J_m)^2}.
\end{align}

\begin{figure}
\includegraphics[scale=0.8]{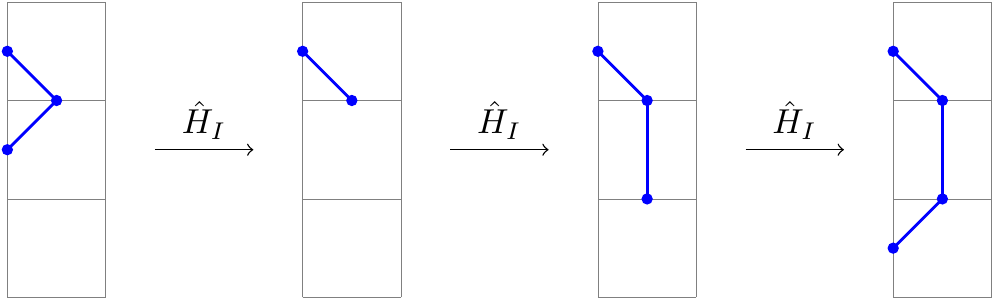}

\caption{Pictorial representation of the tunnelling process of the boundary theory induced by the magnetic field $\hat{H}_I$. For this schemes it is also possible to get the leading order in the same process at distance $R$, Eq.~\eqref{eq:shiftR}.}
\end{figure}
%
%{\color{blue} Maybe we should account for factor 2: $\quad \dfrac{2(h_x)^3}{(-2J_m)^2} $ because of degenerate states as in paper. Should think about it.}
Using the same technique, it is possible to evaluate the energy shift for more "long-range" transitions like $\ket{R=0} \rightarrow \ket{R}$ of which the energy shift at the leading order is 
\begin{equation}\label{eq:shiftR}
\delta E_R = \frac{(h_x)^{R+2}}{(-2J_m)^{R+1}}.
\end{equation}

\section{Dispersion relation of the boundary theory}
We want to study the dispersion relation of the boundary theory. Excitations of this theory are composed by $\sigma_i^x$ operators. We can then use as an effective Hamiltonian the following spin Hamiltonian
\begin{equation}\label{eq:effham}
\hat{H}_{\text{eff}} = \sum_{i,j} J_{i,j}\hat{\sigma}^x_i \hat{\sigma}^x_j.
\end{equation}
As in Ref.~\cite{Yu2013}, $J_{i,j}$ is the probability of transitions connecting two spins on the boundary at a distance $R=\vert i -j \vert$ given in Eq.~\eqref{eq:shiftR}. The physics of long-range coupling in the Ising model can lead to extremely interesting effects both in equilibrium~\cite{Belgio2018, tagliacozzo2012} and out-of-equilibrium~\cite{Cevolani2015,Cevolani2016,Cevolani2017} physics. The behavior of the KTC and the stability of topological order in the presence of a magnetic field was already studied in Ref.~\cite{wootton2011bringing, trebst2007breakdown}. There, the authors derived an effective description for the \textit{bulk} Hamiltonian under the perturbation of generic magnetic field. In our set up the magnetic field is chosen to be small enough to do not perturb the bulk but just the \textit{boundary} degrees of freedom.\\
We can use the translational invariance of the system to write the dispersion relation of the excitations in Fourier space. We transform our Hamiltonian in Fourier space
\begin{equation}
 \sigma_i^x=\frac{1}{\sqrt{\mathcal{L}}}\sum_k e^{-\imath  k i}\sigma_k^x.
\end{equation}
We can then plug in this expansion into Eq.~\eqref{eq:effham} to get:
\begin{equation}
 \mathcal{H}_{\textrm{eff}}=\frac{1}{\mathcal{L}}\sum_{i\,R}\sum_{k\,q} J_R e^{\imath \left(k+q\right)i}e^{\imath qR}\sigma_k^x\sigma_q^x,
\end{equation}
where $\mathcal{L}$ is the length of the boundary of our two--dimensional system. The summation over $i$ can be performed and it gives a delta function which enforces the momentum conservation
\begin{equation}
 \mathcal{H}_{\textrm{eff}}=\sum_R \sum_k J_Re^{\imath k R} \sigma_k^x \sigma_{-k}^x.
\end{equation}
The previous relation can be used to identify the dispersion relation:
\begin{equation}
 \epsilon_k=\sum_R J_R e^{\imath R k}.
\end{equation}
The realness of the dispersion relation is ensured by the fact that $J_R$ is symmetric under $R\rightarrow -R$
\begin{equation}
\sum_R J_R e^{\imath k R} = 2\Re \sum_{R>0} J_R e^{\imath k}.
\end{equation}
We can then use the Eq.~\eqref{eq:shiftR}
\begin{equation}
 J_R=\dfrac{(h_x)^{R+2}}{(-2J_m)^{R+1}}=\frac{h_x}{\left(-2J_m\right)^2}\left(\frac{h_x}{-2J_m}\right)^R.
\end{equation}
\begin{figure}[t]
\includegraphics[width=\columnwidth]{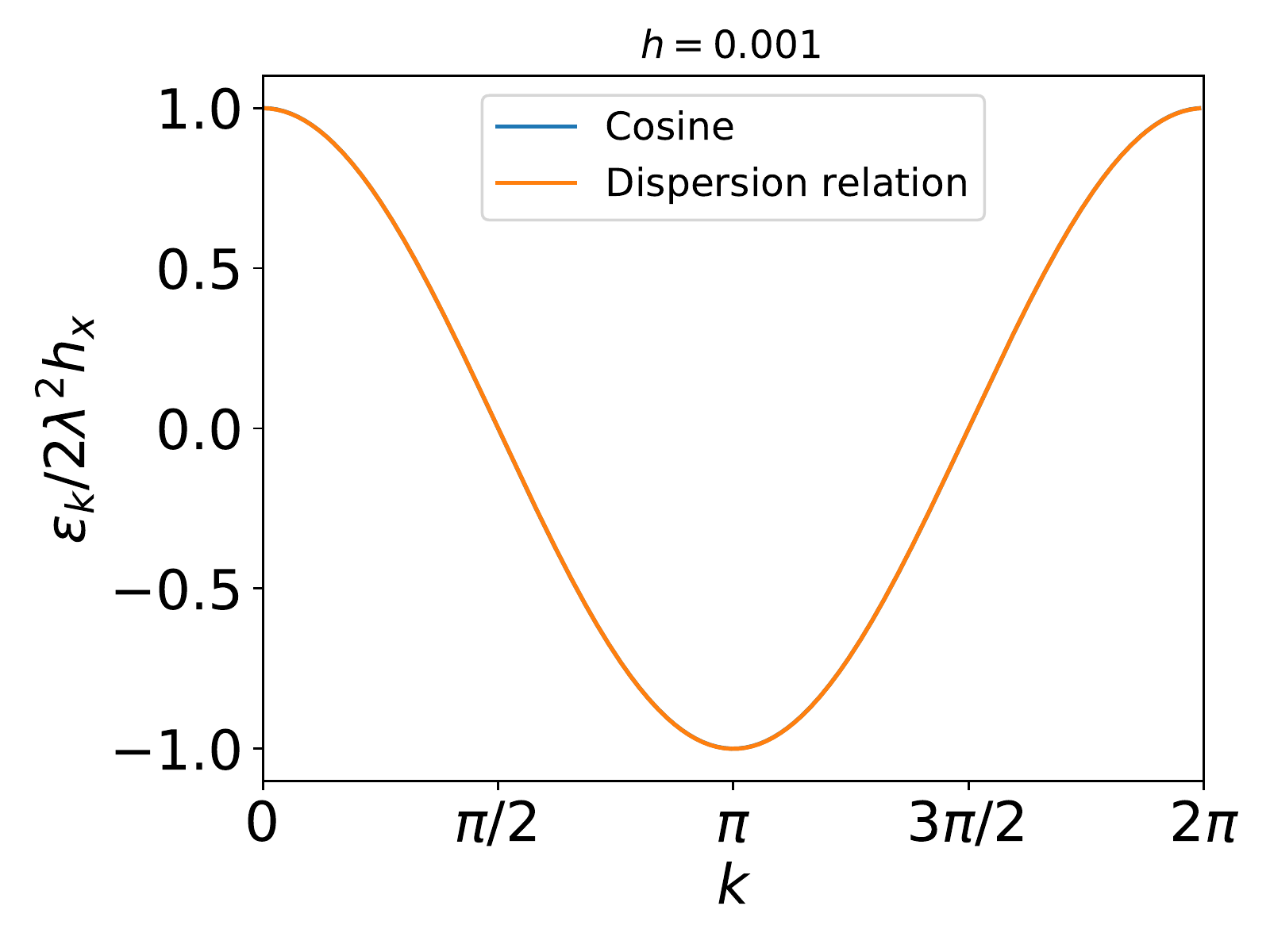}
\includegraphics[width=\columnwidth]{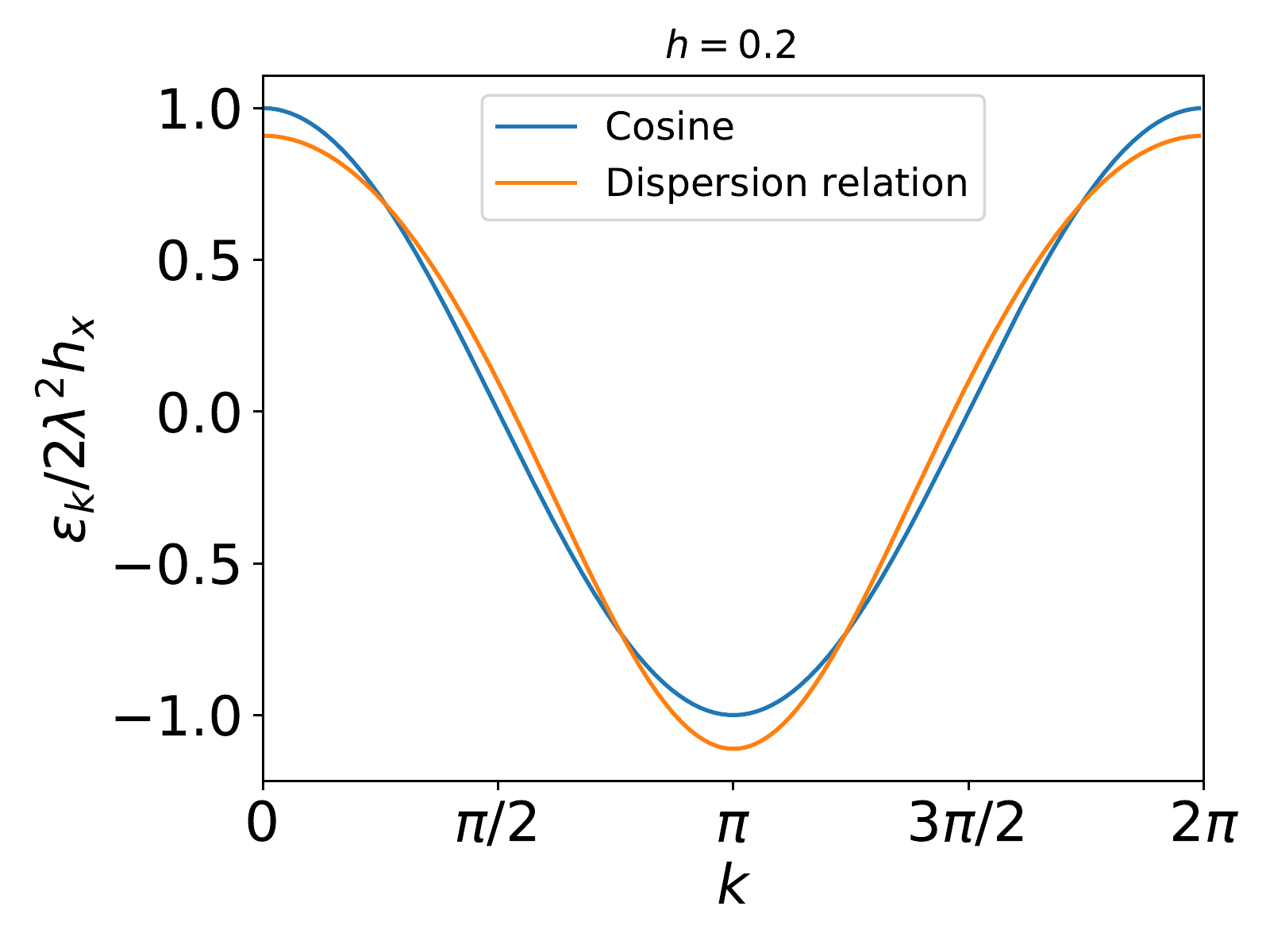}
\caption{\label{fig:disprel}{\footnotesize The dispersion relation of the boundary theory for different values of $\lambda$ compared to a pure cosine. In order to have a better scale, we traced the dispersion relations in unit of $2h_x\lambda^2$. In the upper panel: dispersion relation for $h=0.01$ and $J_m=1$ compared to a pure cosine. It is possible to see that the difference between these two is almost zero, meaning that our theory can be understood using a nearest-neighbour model. In the lower panel we have the same dispersion relation for $h=0.2$ and $j_m=1$, $\lambda=-0.1$, compared again to a cosine function. The deviation from a pure nearest--neighbour hopping are more evident in this case and a more "long-range" physics is taking place.}}
\end{figure}
We can write the dispersion relation using the previous expression and the fact that the sum over $R$ runs from $R=1$ to $R=\mathcal{L}/2$ which is the maximal distance on the boundary.\\
The summation can be performed analytically using the $\lambda=h_x/\left(-2J_m\right)$ as a short-hand notation for the perturbation parameter. We get:
\begin{equation}
\epsilon_k = 2h_x \lambda\Re \sum_{R=1}^{R=\mathcal{L}/2} \left(e^{\imath k} \lambda \right)^R=2h_x \lambda\Re \frac{\lambda e^{\imath k}-\left(\lambda e^{\imath k}\right)^{\frac{\mathcal{L}}{2}+1}}{1-\lambda e^{\imath k}},
\end{equation}
the dispersion relation can then be written in the form:
\begin{equation}
\epsilon_k = 2h_x \lambda^2\frac{\cos\left( k \right)-\lambda +\lambda^{\frac{\mathcal{L}}{2}+1}\cos\left(\frac{ k\mathcal{L}}{2} \right)-\lambda^{\frac{\mathcal{L}}{2}}\cos\left( \frac{k(\mathcal{L}+2)}{2} \right)}{1+\lambda^2-2\lambda\cos\left( k\right)}.
\end{equation}
Taking $\mathcal{L}$ sufficiently large and the parameter $\lambda$ sufficiently small, we can evaluate the previous equation in the thermodynamic limit with respect to the border:
\begin{equation}
\epsilon_k = 2h_x \lambda^2 \frac{\cos\left(k\right)-\lambda}{1+\lambda^2-2\lambda\cos\left(k\right)}.
\end{equation} 
An important approximation can be done in the last expression, if $\lambda$ is sufficiently small, the dominant hopping is the nearest--neighbour hopping, in this case we recover a cosine band
\begin{equation}
\epsilon_k \approx 2h_x \lambda^2 \cos\left(k\right),
\end{equation}
for values of $\lambda$ not negligible. It is possible to see deviations, meaning that the couplings between more distant sites become important, see Fig.~\ref{fig:disprel}

\section{Quantum computation with the KTC with open boundary conditions}
In the end, we also want to discuss the question of whether is it possible to use  KTC with OBC for quantum computation. This question is important to discuss as far as this is of the use for the original KTC with periodic boundary conditions and one should understand if and how does the situation changes when one moves to open boundaries. We can start from the considerations on the standard KTC with PBC, where the information is stored in two topologically protected ground states, winding along the two non--trivial paths of the torus. The memory of such a system is protected by an energy gap. The only error that cannot be detected is the one that does not increase energy, meaning that it has to convert one ground state to another one. Since the ground states are paths along the two main directions of the torus, this requires an exponentially large number of ordered spin-flips. Assuming a constant probability to create an error in the system with one spin--flip, the probability to corrupt the information is exponentially small in the size of the system.\\ 
In our system, the degrees of freedom of the ground states are located at the boundary, given by the $\mathcal{W}_i$ operators. Trivially, one could think about using such operators to store the information one would like to protect. Even though one can naively think that the high complexity and degeneracy of the ground state can make the memory more stable to errors, thus it appears not to be the case. The fundamental reason is that such states are not protected by the topology any more. In fact, to corrupt the information written in our memory, it is sufficient to modify just $3$ spins that correspond to creation or destruction of a $\mathcal{W}_i$ operator. This probability is constant and it does not decrease with the system size as in the PBC case. \\
Even though, as was seen in the previous section, with the presence of external magnetic field, the edge states are no longer gapless, this still does not make the memory robust. The degenerate ground state splits (similar to Zeeman effect) and acquire an energy gap. One should remember though that the gap is still small due to the fact that it is proportional to magnetic field which is weak. Generally, the gap decreases exponentially with the number of spins excited on the edge. This means that the number of the edge states we can use as q--bits and hence the capacitance of our memory is restricted from the requirement of the gap being sufficiently big. Or, in other words, the level of protection we want to achieve restricts the capacity of the memory we can have.  This is, however, as was said before has nothing to do with the topological protection that the memory has in the KTC with PBC. \\
A non-trivial topology can anyway be induced in the system using holes~\cite{Brown2017} and defects~\cite{Brown2013}. Information can then be stored in spin-flips around these points being topologically protected. Planar codes can also be used to do proper quantum computation~\cite{bombin2007optimal,bombin2007homological,nigg2014quantum} instead of simple storage of information.
Even if our model cannot be directly used as a quantum memory, its utility is clear from the mathematical structure of Eq.~\eqref{eq:belgroundstate}. This expression can be used to account analytically for the effect of boundary states on the physical quantities and their response to external perturbations.

\section{Conclusions}

In this paper, we studied the Kitaev Toric Code with particular open boundary conditions. These conditions restrict any exchange of the excitations between the bulk and the boundary, which are then completely decoupled. We wrote down explicitly the expression for the ground state highlighting the separation between the degrees of freedom in the bulk and the ones localised on the boundary (see Eq.~\eqref{eq:belgroundstate}). The presence of the edge modes in open--boundary systems is expected, but our expressions reveal their mathematical form which is not always known. Moreover, the degrees of freedom are located on the boundary but the states themselves traverse deeply into the bulk of the system because of the different operators content of the bulk and the boundary.\\
We demonstrated that the spatial distribution of the degrees of freedom influences important quantities such as the bipartite entanglement entropy. In fact, the difference between open and periodic boundary conditions can be seen only if the bipartition cuts the boundary. There, the presence of these degrees of freedom gives a positive contribution to the leading order which has an area law behaviour.\\
Finally, we studied the response of our system to an external perturbation, namely a weak magnetic field. In this situation, the degrees of freedom at the boundary acquire a dispersion relation and they can be described by a generalised Ising model. Using perturbation theory, we calculated the dispersion relation of the model in this limit. Due to the perturbation, the edge modes cannot enter deep into the bulk, but they are mainly confined to the region close to the boundary and consequently take the shortest path between two points.

\section{Acknowledgements}
We thank Kai P. Schmidt and Roman Orus for very helpful discussion. We thank S. Paeckel and K. Harms for useful comments on our manuscript.

\newpage

%\bibliographystyle{apsrev4-1}
%\bibliography{bibliography}

\end{document}